\documentclass[12pt,english]{article}
\usepackage[affil-it]{authblk}

\usepackage[]{geometry} 
\usepackage{babel}
\usepackage{graphicx} 
\usepackage{pict2e}
\usepackage{float}
\usepackage{cite}
\usepackage{eqnarray,amsmath}

\def\sgn{\mathop{\rm sgn}\nolimits} 

\title{An attractor neural network architecture with an ultra high information capacity: numerical results}
\author{Alireza Alemi }
\affil{\small{Group for Neural Theory, \'Ecole Normale Sup\'erieure,\\ 29 rue d'Ulm - Paris 75005, France}\\{{email: alireza.alemi@ens.fr}}
}
\date{} 
\begin{document}


\maketitle
\begin{abstract}
Attractor neural network is an important theoretical scenario for modeling memory function in the hippocampus and in the cortex. In these models, memories are stored in the plastic recurrent connections of neural populations in the form of ``attractor states''. The maximal information capacity for conventional abstract attractor networks with unconstrained connections is 2 bits/synapse. However, an unconstrained synapse has the capacity to store infinite amount of bits in a noiseless theoretical scenario: a capacity that conventional attractor networks cannot achieve. 

Here, I propose a hierarchical attractor network that can achieve an ultra high information capacity. The network has two layers: a visible layer with $N_v$ neurons, and a hidden layer with $N_h$ neurons. The visible-to-hidden connections are set at random and kept fixed during the training phase, in which the memory patterns are stored as fixed-points of the network dynamics. The hidden-to-visible connections, initially normally distributed, are learned via a local, online learning rule called the three-threshold learning rule and there is no within-layer connections. 

The results of simulations suggested that the maximal information capacity grows exponentially with the expansion ratio $N_h/N_v$.  As a first order approximation to understand the mechanism providing the high capacity, I simulated a naive mean-field approximation (nMFA) of the network. The exponential increase was captured by the nMFA, revealing that a key underlying factor is the correlation between the hidden and the visible units. Additionally, it was observed that, at maximal capacity, the degree of symmetry of the connectivity between the hidden and the visible neurons increases with the expansion ratio. These results highlight the role of hierarchical architecture in remarkably increasing the performance of information storage in attractor networks.
 \end{abstract}
\section{Introduction}

An auto-associative or attractor neural network is a popular theoretical framework for modeling long-term memory and cortical information processing in the brain \cite{hopfield82,amit89,hertz91,lansner09}. Originally conceptualized by HebbÕs learning and neural assembly \cite{hebb49}, the framework links single neuron activity to the storage of a memory concept as an emergent cognitive process: a memory concept is distributed over a collection of neurons whose synapses are modified according to the Hebbian rule to store the concept. Technically speaking, the memories are (point) attractor states or stable fixed points of the network dynamics. Upon activation of a cue partially correlated with a stored memory, the network dynamics evolve towards the stored memory; therefore, the network ÔrecallsÕ the memory. The set of states from which the network evolves towards an attractor state is called the basin of attraction of the attractor. The size of basin of attraction determines how robust a memory is against noise.

In electrophysiological recordings in awake monkeys that performed delayed response tasks, some neurons exhibited selective persistent activity during the delay period which have been considered as neural correlates of working memory. Such activities are consistent with the attractor network scenario. Candidate structures for attractor neural networks are horizontal connections of pyramidal cells in layers 2/3 and 5 of higher-order association cortex \cite{rolls06} and the recurrent CA3-CA3 connections in the hippocampus \cite{rolls07}.

Two measures of performance of attractor neural networks are: the storage capacity defined as the number of patterns that can be stored in a network of $N$ neurons; and the information capacity defined as the total entropy of binary patterns that can be stored per synapse. The storage capacity of the Hopfield model with uncorrelated patterns and dense coding is $0.138N$ \cite{amit85}, while the maximal capacity (or Gardner bound) of the recurrent network in the dense coding is $2N$ \cite{gardner88} (or in this case equivalently $2$ bits/synapse).

In a noiseless information theoretic limit for an ideal observer who has direct access to synapses corresponding to a memory trace, the information capacity is $log_2(L)$ bits/synapse for a synapse with $L$ discrete states and this limit diverges for continuous synapse, i.e. continuous synapses theoretically have infinite information capacity for storage.  Yet, the Gardner bound (which is for recurrent networks of unbounded, continuous synapses) is restricted to only $2$ bits/synapse. This means that the conventional attractor networks cannot fundamentally achieve this theoretical infinite capacity. It is worth noting that neuronal networks, under evolutionary pressure, may need to use their resources optimally. Therefore, to store memories optimally, they may need to operate close to the maximal capacity. In fact, studying properties of an optimal network for storage has provided a theory of synaptic weight distribution that matches experimental one in the Purkinje cells in the cerebellum \cite{brunel04}.

Given the current state of capacity in attractor neural networks,  questions that may arise are: does a structure exist for attractor memory networks that brings their capacity close to the ideal observer information capacity?
 Could a local learning rule in such a network store information close to the theoretical limit?

\section{The Model}
The network structure is made of two layers: a visible layer and a hidden layer. The connections between layers are fully-connected without lateral connections within layers. More specifically, each hidden neuron $Y_i$ (with the index $i\in\{1, 2, ..., N_h\}$) receives inputs, weighted by $v_{ij}$, from the visible units $X_j$ with the index $j\in\{1,2,...,N_v\}$, where $N_h=\lambda N_v$ (see Fig. 1A). A visible neuron $X_j$ receives inputs from the hidden units weighted by the synaptic weights $w_{ji}$. Aiming for a proof of concept here, I considered an abstract, deterministic neuron model with $\pm 1$ output. The dynamics of the networks at time $t$ (synchronous update), in the absence of external inputs (patterns) follow as

\begin{eqnarray}
	Y_{i}^{t} = \sgn \Big(\sum_{j=1}^{N_v} v_{ij}X_{j}^{t-1}\Big) \label{eq:1}\\
	X_{j}^{t} = \sgn \Big(\sum_{i=1}^{N_h} w_{ji}Y_{i}^t\Big) ,\label{eq:2}
\end{eqnarray}
 where $\sgn$ stands for the Sign function: $\sgn(z)=+1$ if $z>0$ and $-1$ otherwise.

The goal of the model is to store a set of $M$ uncorrelated, binary ($\pm1$) patterns $\{\vec{\xi}^\mu\}$ (where $\mu\in\{1,2,...,M\}$) as fixed-points of the dynamics of the network i.e. for each $j$ and $\mu$ the following equations must hold:
\begin{equation}
\xi_j^{\mu} = \sgn \bigg( \sum_i w_{ji} \sgn \Big(\sum_l v_{il} \xi_l^{\mu}\Big)\bigg).
\label{eq:fixedpoint}
\end{equation} The binary random variables $\xi_j^\mu$ are independent from each other and the probability of $\xi_j^\mu$ becoming active is 0.5 (hence the entropy of each entry of the pattern matrix is one), thus being at the dense regime. The fixed feedforward weights $v_{ij}$ 
 are sampled from a Gaussian distribution with mean zero and standard deviation one [denoted by $\mathcal{G}(0,1)$], ensuring the hidden units work at the dense regime as well. On the other hand, the plastic feedforward weights $w_{ji}$ were modified during the fixed-point learning process.

In this training phase, the number of patterns scales as $M=\alpha N_h$. The quantity $\alpha$, called the storage capacity ratio, coincides, in the dense regime, with the information capacity which is the entropy of pattern matrix  divided by the number of plastic connections: $I= {M  N_v . 1 \over N_h N_v}={M \over N_h}=\alpha$.  
Instead of the perceptron learning rule (PLR) which uses an external `error signal' (i.e. the difference between actual output and desired output), I used the three-threshold learning rule (3TLR) where the error signal is computed internally by using three threshold set on the total synaptic inputs of a neuron\cite{alemi15}. 

In the learning phase, the patterns are presented sequentially in random order. During the presentation
of a pattern $\vec{\xi}^\mu$ each neuron $X_j$ receives an external binary input $\hat\xi_i^\mu = \chi \xi_i^\mu$, where $\chi$ denotes the strength of the external inputs and is parameterized as $\chi = \gamma \sqrt{N_v}$ and recurrent inputs from the hidden neurons. The parameter $\gamma$ is set such that the strong external inputs sets the states of the visible units, not allowing them to be affected by the recurrent inputs, therefore matching the performance of the learning to the PLR. Once the external inputs $\vec{\hat\xi}^\mu$ at time $t$ is presented, the visible neurons are clamped to their desired states, i.e.  $\vec{X^t}=\vec{\xi}^\mu$,  and they update the states of the hidden neurons $Y_i^{t+1}$ according to Eq.~\ref{eq:1}, keeping them fixed as long as the visible neurons are clamped. The new states of the hidden neurons update the $x_j^{t+1}$'s, the total synaptic input to the visible units $j$, as

\begin{equation}
x_j^{t+1}=\sum_{i=1}^{N_h} w_{ji}^tY_{i}^{t+1}+\gamma\sqrt{N_v}\xi_i^{\mu}
\label{eq:localfield}
\end{equation}

but do not change $X_j^{t+1} = \sgn(x_j^{t+1} ) = X_j^t$ since $\gamma$ is large enough (see\cite{alemi15} for more details). Once the $x_j^{t+1}$'s are computed, they are used to update the $w_{ji}$'s according to the local rule (3TLR): 

\begin{equation}
	w_{ji}^{t+1}=
	\begin{cases}
		w_{ji}^t - \eta Y_i^{t+1}, & \text{if } -{\gamma\sqrt{N_v}\over{2}} < x_j^{t+1} < 0 \\

		w_{ji}^t + \eta Y_i^{t+1}, & \text{if ~~~~~~~} 0 ~< ~x_j^{t+1} < {\gamma\sqrt{N_v}\over{2}}\\

		w_{ji}^t ,&  \text{otherwise, }
	\end{cases}\label{eq:3TLR}
\end{equation}
where $\eta=0.001$ is the learning rate. After this weight update, the pattern $\mu$ is removed, another pattern is presented, and the above procedure continues.

The set of patterns are presented to the network for a number of times. After some number of presentations, it was checked whether the patterns are learned i.e. whether the patterns $\{\vec{\xi}^\mu\}$ are the fixed points of the network dynamics (Eq.~\ref{eq:fixedpoint}). A hard limit was imposed on the number of pattern presentations which was 5000 iterations. If after this maximum number of presentations, the patterns were not learned, the simulation was stopped, and the storage of the pattern set was considered unsuccessful.

\begin{figure}
\begin{center}
\includegraphics[scale=0.9]{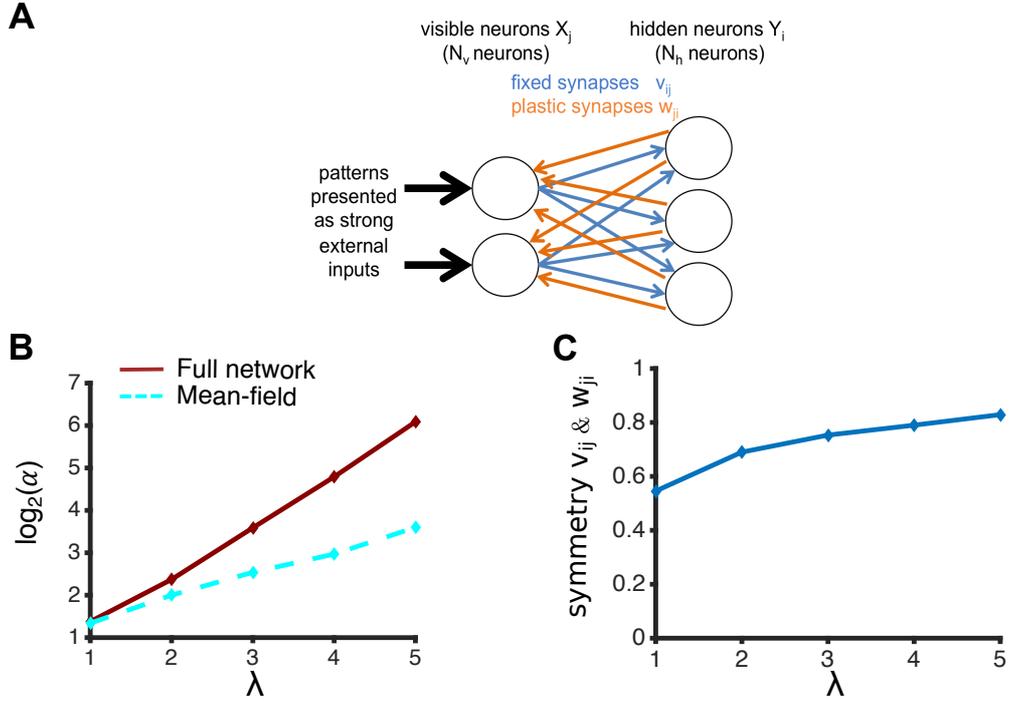}
\end{center}
\label{fig:1}
\vspace{-2em}
\caption{The architecture of the network and its main results. \textbf{A}. The network architecture which consists of two layers. The visible-to-hidden weights $v_{ij}$ are sampled from a Gaussian distribution and kept fixed during learning. The hidden-to-visible weights $w_{ji}$ are trained to store the  patterns while patterns are presented as strong external inputs. \textbf{B}. The logarithm of capacity as a function of expansion ratio $\lambda=N_h/N_v$ for the full network and the nMFA averaged over 10 random iterations. Both of the traces of close to straight lines which suggests that the capacity increases exponentially as a function of $\lambda$  \textbf{C}. The mean degree of symmetry between $v_{ij}$ and $w_{ji}$ increases with expansion ratio $\lambda$.}
\end{figure}

\section{Results}
The numerical results are depicted in Fig. 1B and Fig. 1C showing logarithm of maximal capacity and the degree of symmetry of connections as a function of expansion ratio $\lambda$ for $N_v=100$ obtained for 10 random seeds. For each seed we obtained wether the storage was successful or not. In Fig. 1B, the dark red line shows $\log_2(\alpha)$ at which the probability of successful storage is 0.5. This tends to be a straight line suggesting that information capacity grows exponentially with $\lambda=N_h/N_v$. Such capacities significantly surpass the 2 bits/synapse information capacity in conventional attractor networks \cite{gardner88}. Additionally, the degree of symmetry of the weights between the visible and the hidden units, measured by the correlation coefficient between $v_{ij}$ and $w_{ji}$, at maximal capacity increases with $\lambda$ (Fig. 1C).

As a first attempt to understand the underlying mechanism of providing such a high capacity, I considered the inputs from one of the visible neurons (say $X_1$ without loss of generality) to the hidden neurons as the desired signal and the inputs from other visible, i.e. $\sum_{j=2}^{N_v}v_{ij}X_j$,  as a \textit{quenched} Gaussian noise $\mathcal{G}(0,\sqrt{N_v-1})$. The inputs from $X_1$ to $Y_j$ cause a correlation between $X_1$ and $Y_j$ which can increase the capacity for learning the weights $w_{j1}$.  This approximation is called a naive mean field approximation (nMFA) since the conditional probability distributions of $\vec{Y } \vert X_1$ factorizes: $P(\vec{Y}  \vert X_1)=\prod_i{P(Y_i\vert X_1)}$. The term quenched noise here means that once the noise distribution is sampled for a hidden neuron and pattern index $\mu$, the samples become associated with the pattern index and the hidden neuron and they do not change during the iterations of learning dynamics.
The capacity of the nMFA increases exponentially with $\lambda$ (though with a lower exponent), therefore capturing the essence of the phenomenon (Fig. 1B; cyan dashed line). 

\section{Discussion}

Although the attractor-memory framework provides a mechanistic explanation for storage and retrieval of long-term memory, its performance is far from ideal, i.e. the maximal information per synapse is far from that of the ideal observer.
This proposal puts forward the hypothesis that attractor states can be built in
a network with hierarchical structure where a hidden layer with large expansion ratio can
greatly increase the information capacity. Hidden neurons provide a random projection (basis functions), a new internal representation allowing the network to store more efficiently attractor-memories through synapses. The nMFA shows that weak correlation between hidden units and visible units is a key factor providing such a high capacity. 
A future work will include analytical results for the nMFA which acts as a lower bound for the capacity of the original network. The current results concern the zero size of basin of attraction. I will improve on that by increasing the basin using robustness parameters as allowed by the 3TLR \cite{alemi15}.
Here for the sake of simplicity the binary neuron model was used. It would also be very interesting to extend this work to see how this architecture can be implemented using spiking neuron models.

The current brief article serves to identify and define the research problem of capacity. It addresses the issue by proposing a hierarchical architecture and testing it with numerical simulations. It also attempts to give a first approximation of the model such that it would provide an understanding of the underlying cause of such a high capacity. The next step is to provide an analytical calculation for capacity of nMFA to fully determine the lower bound for the maximal capacity. It would be interesting to think about a way to analytically calculate the maximal capacity of the full network.

\section*{Acknowledgement}

 I would like to thank Nicolas Brunel and Sophie Deneve for their feedbacks and comments on an earlier version of the text. I am grateful to Sahar Pirmoradian and Carlo Baldassi for fruitful discussions and comments.

\bibliography{highcapacity}

\end{document}